\documentclass[aps,prl,reprint,twocolumn,floats,showpacs,letterpaper,floatfix]{revtex4}
\usepackage[T2A]{fontenc}
\usepackage[latin9]{inputenc}
\usepackage[active]{srcltx}
\usepackage{graphicx}
\usepackage{esint}
\makeatletter

\providecommand{\tabularnewline}{\\}

\makeatother

\begin{document}

\title{How the quantum ring shape influences its energy spectrum}
\author{V.~A.~Roudnev}
\author{A.~M.~Puchkov}
\author{A.~V.~Kozhedub}
\affiliation{Physics Department, St-Petersburg State University, Russia}
\begin{abstract}
We propose a new approach for modeling the quantum ring single particle
energy spectrum. The approach is based on separation of variables
in the Schr\"odinger equation in oblate spheroidal coordinates. We consider
a model of a spheroidal quantum ring with infinite walls. Our simple
model allowed us to study the spectra for quantum rings of different
shapes. The spectrum is calculated for the ground and several excited
states and the ring shape dependence of the spectrum is demonstrated.
The spectrum can demonstrate parabolic or non-parabolic dependence
on the magnetic quantum number for different shapes of the ring profile.
\end{abstract}
\pacs{73.21.-b, 73.22.Dj, 71.15.-m}
\maketitle

\section{Introduction}

Quantum dots and quantum rings are becoming irreplaceable components
of modern electronics due to the arising technological opportunity
to control their spectral properties accurately. Such structures as
multiple concentric nano-rings, rings around a quantum dot and many
other complex nano-objects are been fabricated on the base of droplet
epitaxy \cite{SelfOrgQRFabrication, DotRingFabrication, MultiRingsFabrication}.
Fabrication methods for creating regular two- and three-dimensional
structures of nano-rings are also being developed on the base of nanospherical
lithography \cite{RingArrays1, 3DNanoArrays, ColloidalTemplate}.

Despite an obvious progress in fabricating nano-objects, however,
the researchers still meet a challenge of their efficient theoretical
description. Even for single-particle states, we have an alternative:
either to restrict ourselves to oversimplified one-dimensional models
\cite{chaplik}, or to resort to computationally costly calculations
for taking into account the three-dimensional structure of the systems
\cite{Voskoboynikov, filikhin}. In this work we suggest an approach
which on the one hand allows us to treat quite complex three-dimensional
ring-shaped nano-structures, and on the other hand to reduce the computational
cost of the model by exact separation of variables in oblate spheroidal
coordinates.

The major goal of this work is to demonstrate suitability of oblate
spheroidal coordinates for describing three-dimensional nano-scale
quantum rings using a simple model of a particle confined to a ring-like
structure by an infinite potential wall. This approach not only allowed
us to develop a computationally simple scheme for ring spectrum calculations,
but also to study the dependence of such spectra on the shape of the
ring.

It is worth mentioning that J.~Even and S.~Loualiche \cite{Even}
have also considered a model of a quantum ring bounded by an infinite
potential wall. In their case, the boundary of the ring is formed
by circular paraboloids and the separation of variables is performed
in parabolic coordinates. The advantage of our approach over their
is that it allows us to vary the shape
of the ring while fixing the volume and the characteristic size $R$
of the ring, which, basically, has made our study possible. In our
approach it is also possible to model a flat substrate directly without
solving an auxiliary problem for a symmetric ring and post-selecting
the solutions that fit the required boundary condition.

It is well known, that the two types of the spheroidal coordinates
-- prolate and oblate -- both admit separation of variables for the
Schr\"odinger equation \cite{kom_pon_sla, morse_fesh}. Even though there
are only a few known quantum problems with exact separation of variables,
it is the prolate spheroidal coordinates that are preferably used
in hundreds of published researches. The oblate spheroidal coordinates,
however, have been used rather rarely in quantum mechanical calculations.
There are only a few such papers known to the authors. The most famous
example is, probably, the the work of Rainwater \cite{Rainwater},
where the model of a spheroidal infinitely deep potential well made
it possible to explain magnetic moments of many nuclei based on the
behavior of an unpaired nucleon. In this letter we demonstrate how
the use of oblate spheroidal coordinates allowed us to make some curious
observations on the variations of quantum ring spectra as the ring
shape changes.

\begin{figure}[t!]
\centering \includegraphics[width=0.43\textwidth]{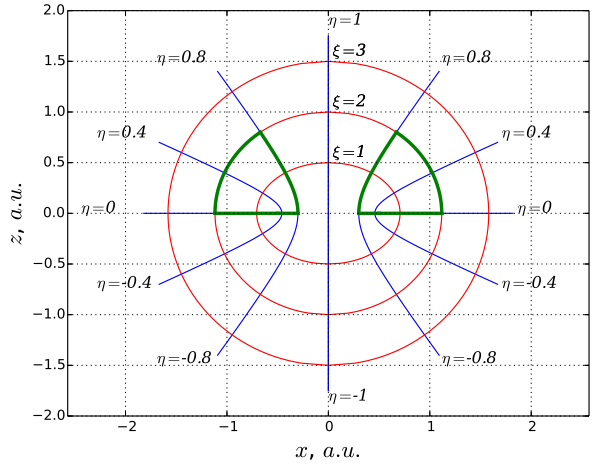}
\caption{Oblate spheroidal coordinate surfaces projection to $(x,z)$-plane
with $z$ as the symmetry axis. The bold contour corresponds to a
cross section of the quantum ring boundary. The three-dimensional
shape of the ring is given by rotating around the $z$ axis by the
angle $0\leq\theta\leq2\pi$. (Color online).}

\label{OSC}
\end{figure}

\begin{figure}[t!]
\centering \includegraphics[width=0.43\textwidth]{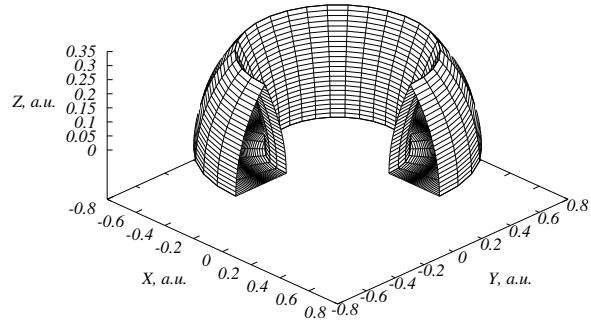}
\caption{Isometric projection of a quantum ring, a rectangular segment is taken
off to demonstrate the cross section.}
\label{Ring}
\end{figure}

We employ a very simplified model of a single particle in a quantum
ring. Assuming a very sharp transition between inner and outer regions
of the quantum well we can neglect the effective mass inhomogeneity.
As we are interested in qualitative properties of the particle confined
in a ring, we use natural units (n.u.) of energy such that the Schr\"odinger
equation for a free particle takes the following form
\begin{equation}
\frac{1}{2}\Delta\Psi+E\Psi=0\,.
\label{a1}
\end{equation}
We introduce oblate spheroidal coordinates$(\xi,\eta,\varphi)$
\begin{equation}
\begin{array}{l}
x=\frac{R}{2}\sqrt{(\xi^{2}+1)(1-\eta^{2})}\cos{\varphi},\\
y=\frac{R}{2}\sqrt{(\xi^{2}+1)(1-\eta^{2})}\sin{\varphi},\\
z=\frac{R}{2}\xi\eta,\\
\xi\in[0,\infty),\;\;\;\eta\in[-1,1],\;\;\;\varphi\in[0,2\pi).
\end{array}
\label{a2}
\end{equation}
We shall require the quantum well to be bounded by the coordinate
surfaces $\xi=\xi_{0}$, $\eta=\eta_{0}$ and the plane $\xi=0$, $\eta=0$
which corresponds to a flat substrate. In Fig.~\ref{OSC} we show
a cross section of the coordinate surfaces, and the corresponding
three-dimensional configuration of the ring is shown in Fig.~\ref{Ring}.
Other -- even more complex -- combinations of coordinate surfaces
can also be employed.
By putting zero boundary conditions at the surface of the ring Fig.~\ref{Ring}
we confine the particle inside the ring. Given this boundary conditions
the wave function $\Psi_{j}$ corresponding to the energy $E_{j}$
can be represented as a product
\begin{equation}
\Psi_{j}(\xi,\eta,\varphi;R) = N_{kqm}(R)X_{mk}(\xi;R)Y_{mq}(\eta;R)e^{im\varphi}\,,
\label{a3}
\end{equation}
where the multi-index $j=\lbrace kqm\rbrace$ stands for a set of
quantum numbers $k$, $q$ and $m$ such that $k$ and $q$ give the
number of roots of the corresponding functions in $\xi$ and $\eta$,
while the magnetic quantum number $m$ takes the values of $0,\pm1,\pm2,\dots$.
The normalization constant $N_{kqm}(R)$ can be determined from the
condition
\begin{equation}
\int_{V}\Psi_{kqm}^{\ast}(\xi,\eta,\varphi;R)\Psi_{k'q'm'}(\xi,\eta,\varphi;R)dV=
\delta_{kk'}\delta_{qq'}\delta_{mm'}\,,
\label{a4}
\end{equation}
where
$dV=\frac{R^{3}}{8}(\xi^{2}+\eta^{2})d\xi d\eta d\varphi$
is the volume element in oblate spheroidal coordinates.

Substituting (\ref{a3}) into (\ref{a1}) we obtain a system of ordinary
differential equations \cite{kom_pon_sla}
\begin{equation}
\begin{array}{rcl}
\frac{d}{d\xi}(\xi^{2}+1)\frac{d}{d\xi}X_{mk}(\xi;R) & -\\
-\left[\lambda-p^{2}(\xi^{2}+1)-\frac{m^{2}}{\xi^{2}+1}\right]X_{mk}(\xi;R) & = & 0\,,
\end{array}\label{a5}
\end{equation}
\begin{equation}
\begin{array}{rcl}
\frac{d}{d\eta}(1-\eta^{2})\frac{d}{d\eta}Y_{mq}(\eta;R) & +\\
+\left[\widetilde{\lambda}-p^{2}(1-\eta^{2})-\frac{m^{2}}{1-\eta^{2}}\right]Y_{mq}(\eta;R) & = & 0\,.
\end{array}
\label{a6}
\end{equation}
subject to boundary conditions
$X_{mk}(0;R)=X_{mk}(\xi_{0};R)=0$
and
$Y_{mq}(0;R)=Y_{mq}(\eta_{0};R)=0.$ Here $p_{j}^{2}=\frac{E_{j}R^{2}}{2}\,$
is the energy parameter, $\lambda=\lambda_{mk}^{(\xi)}(p)$ and $\widetilde{\lambda}=\lambda_{mq}^{(\eta)}(p)$
are the separation constants.
The energy spectrum is determined from
the separation constant matching condition
\begin{equation}
\lambda_{mk}^{(\xi)}(p)=\lambda_{mq}^{(\eta)}(p)\,.
\label{a7}
\end{equation}
Obviously, the spectrum scales as an inversed square of the characteristic
ring size $R.$

Consider a set of rings of a fixed volume $V$
\begin{equation}
V=\frac{\pi R^{3}}{4}\int\limits _{0}^{\xi_{0}}\int\limits _{0}^{\eta_{0}}(\xi^{2}+\eta^{2})d\xi d\eta=
\frac{\pi R^{3}}{12}\xi_{0}\eta_{0}(\xi_{0}^{2}+\eta_{0}^{2})\,.
\label{b1}
\end{equation}
Evidently, the parameters of the ring $\xi_{0}$ and $\eta_{0}$ enter
this formula symmetrically. The ranges for $\xi$ and $\eta$, however,
and the corresponding coordinate surfaces are different (Fig.~\ref{OSC}).
We can, thus, fix the volume and the characteristic radius of the
ring in (\ref{b1}) and obtain a relationship between $\xi_{0}$ and
$\eta_{0}$ $\xi_{0}=f(\eta_{0})$ which keeps the volume of the structure
invariant. This way our approach allows us to study the influence
of the ring shape on the structure of its spectrum.

The shape of the ring, however, is not managed by the parameter $\eta_{0}$
alone. Even though the spectrum of the model scales as the inversed
square of the characteristic size of the ring $1/R^{2}$, this scaling
breaks if we keep the volume of the ring fixed. So in order to make
a comprehensive study of the ring shape effects, we should also vary
the volume of the ring or its characteristic size. As the ring volume
scales exactly as $R^{3}$, it is just natural to introduce a dimensionless
parameter $\sigma=\frac{V^{1/3}}{R}$ and use it as the second independent
shape parameter.

\begin{figure}
\begin{tabular}{ccc}
\includegraphics[width=0.14\textwidth]{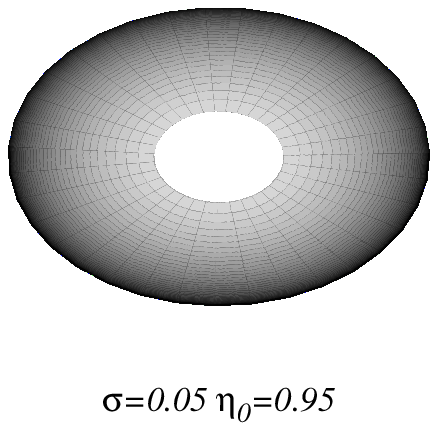} & \includegraphics[width=0.14\textwidth]{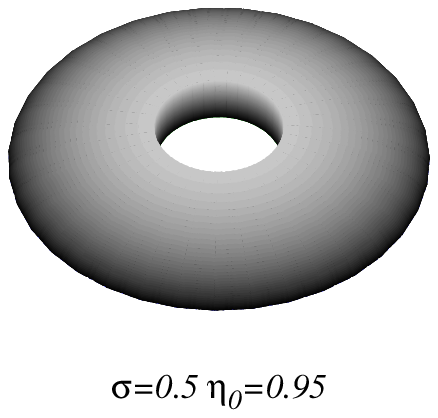} & \includegraphics[width=0.14\textwidth]{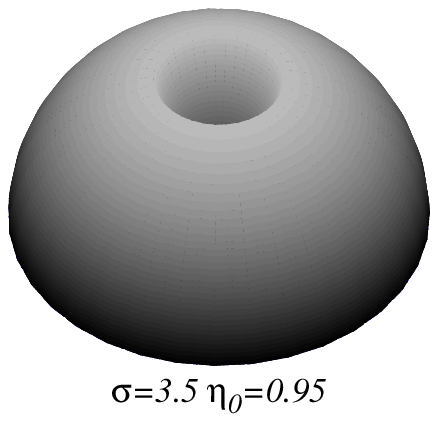}\tabularnewline
\includegraphics[width=0.14\textwidth]{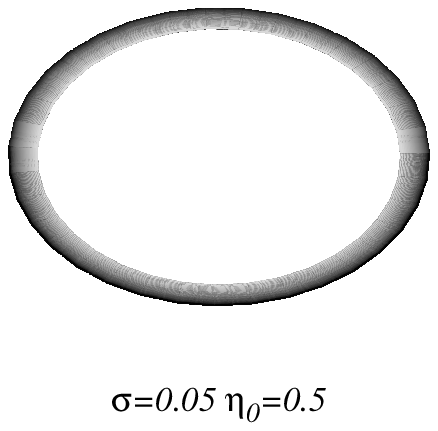} & \includegraphics[width=0.14\textwidth]{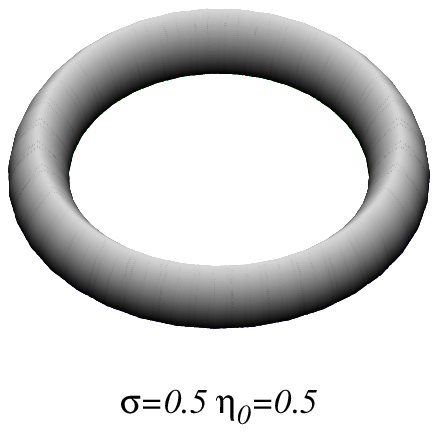} & \includegraphics[width=0.14\textwidth]{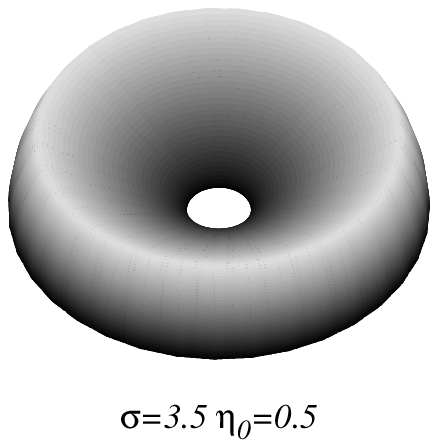}\tabularnewline
\includegraphics[width=0.14\textwidth]{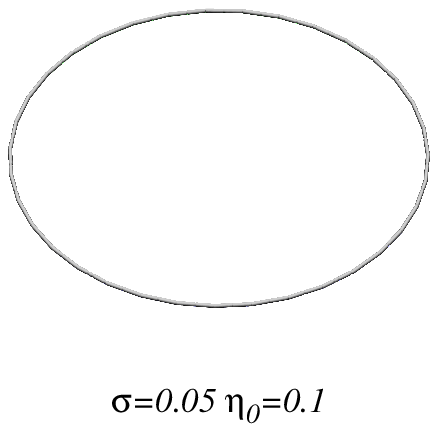} & \includegraphics[width=0.14\textwidth]{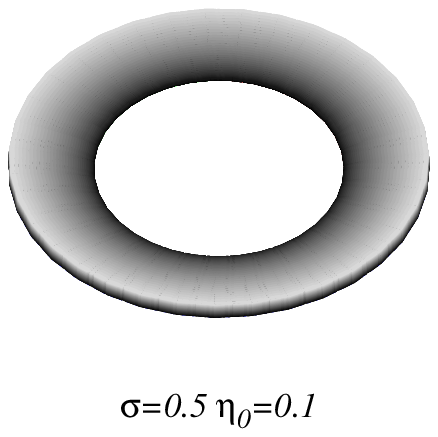} & \includegraphics[width=0.14\textwidth]{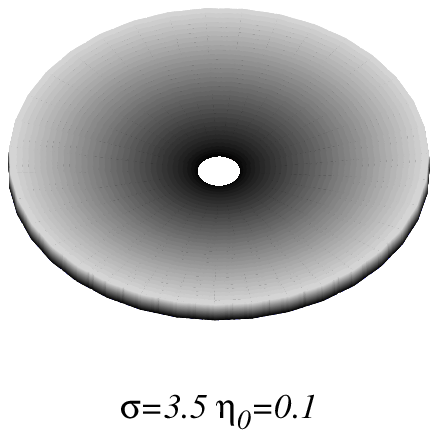}\tabularnewline
\end{tabular}
\caption{Ring shapes for different shape parameters $\eta_{0}$ and $\sigma$.\label{fig:shapes}}
\end{figure}

We illustrate the variations of the ring shape for different values
of shape parameters $\eta_{0}$ and $\sigma$ in Fig.~\ref{fig:shapes}.
We can identify several distinctive cases. For smaller values of $\eta_{0}$
and bigger values of $\sigma$ the ring surface is dominated by the
hyperboloid inside the ring with nearly cylindrical section of the
ellipsoid outside the ring. As $\eta_{0}$ approaches 1 for smaller
$\sigma$ we see the picture reversed: the major part of the boundary
is formed by the ellipsoid outside with a nearly cylindrical section
of the hyperboloid inside the ring. For bigger $\sigma$ the ring
looks like an ellipsoid with a hole. When the both shape parameters
are small the ring looks like a one-dimensional structure.
\begin{figure}
\includegraphics[clip=true,width=0.43\textwidth]{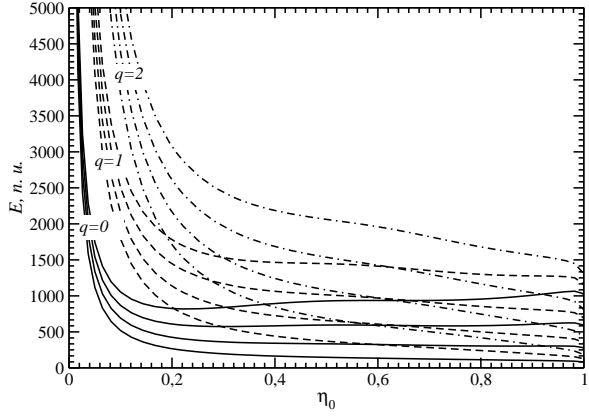} \caption{Energy dependence $E_{kq0}=E_{kq0}(\eta_{0})$ on the shape parameter
$\eta_{0}$ for the ground and the lowest 11 excited states for $\sigma=0.5$, $R=1$. }
\label{terms}
\end{figure}
In Fig.~\ref{terms} we show the energy dependence $E_{j}(\eta_{0})$
on the shape parameter $\eta_{0}$ for 12 lowest eigenstates of the
ring. As $\eta_{0}\rightarrow 0$ the ring is getting flat, and the
energy spectrum starts becoming degenerate in quantum number $k$
as the corresponding degree of freedom contributes less and less to
the total energy. It is also noteworthy that some of the curves $E_{j}(\eta_{0})$
have minima. This newly discovered observation might have some implications
for ring fabrication techniques.

Another, and, probably, more interesting example of the quantum ring
spectrum shape dependence is the study of excitations in magnetic
quantum number $m$. It is usually assumed that the spectrum of angular
excitations in a quantum ring can be described by a simple one-dimensional
model which predicts parabolic behavior of the excited states $E_{m}\propto m^{2}$
\cite{Viefers}.
Our calculations, however, clearly demonstrate essential deviations
from this rather common assumption.

\begin{figure}
\includegraphics[width=0.43\textwidth]{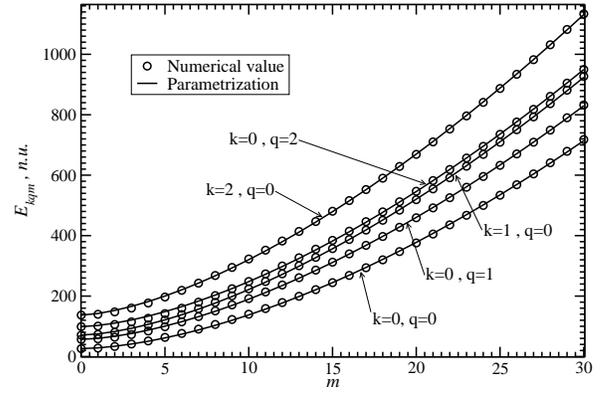}
\caption{Quantum ring excitations and their fits for $V=1/2$, $R=1$ and $\eta_{0}=0.95$.\label{fig:mSpecFits}}
\end{figure}
\begin{table}
\begin{tabular}{|c|c|c|c|}
\hline
$kq$ & $\eta_{0}=0.1$ & $\eta_{0}=0.5$ & $\eta_{0}=0.95$\tabularnewline
\hline
\hline
00 & $200.9+0.16\ m^{1.99}$ & $47.9+0.66\ m^{1.88}$ & $26.1+2.46\ m^{1.66}$\tabularnewline
\hline
01 & $700.8+0.14\ m^{2.00}$ & $127.8+0.50\ m^{1.96}$ & $56.7+3.33\ m^{1.60}$\tabularnewline
\hline
02 & $1491.0+0.14\ m^{2.00}$ & $242.8+0.46\ m^{1.97}$ & $98.9+3.85\ m^{1.59}$\tabularnewline
\hline
10 & $262.4+0.19\ m^{1.99}$ & $100.2+0.79\ m^{1.88}$ & $70.8+3.99\ m^{1.58}$\tabularnewline
\hline
20 & $323.6+0.21\ m^{1.97}$ & $170.9+0.79\ m^{1.91}$ & $137.0+5.31\ m^{1.54}$\tabularnewline
\hline
\end{tabular}
\caption{Parametrization $\epsilon_{kqm}=\epsilon_{kq0}+\alpha_{kq}m^{\beta_{kq}}$
of the spectra for the rings of different shapes for $V=1/2$ and
$R=1$.\label{tab:mSpectra}}
\end{table}

\begin{figure}
\includegraphics[clip=true,width=0.43\textwidth]{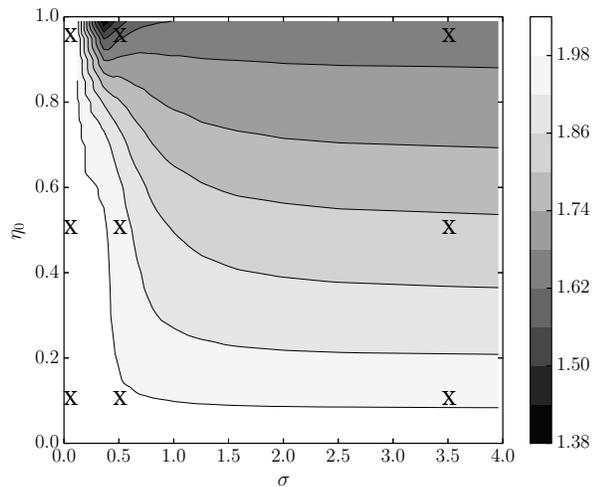}
\caption{The magnetic quantum number excitation spectrum shape characteristic
$\beta_{00}$ for rings of different shapes. The marked points correspond
to the ring configurations in Fig.~\ref{fig:shapes}.\label{fig:spectrumShapeMap}}
\end{figure}

Consider rings of different shapes as shown at Fig.~\ref{fig:shapes}
at a fixed volume and calculate the lowest excitations $E_{kqm}$
for $(kq)=(00), (01), (02), (10), (20)$ and $|m|=0, 1, \ldots,30.$ These
energy levels are smooth functions of the quantum number $m,$ and
their dependence on $m$ is easy to fit with a simple parametrization
$\epsilon_{kqm}=\epsilon_{kq0}+\alpha_{kq}m^{\beta_{kq}}$ as is shown
in Fig.~\ref{fig:mSpecFits}. The one-dimensional ring model corresponds
to $\beta_{kq}\equiv2,$ and the deviations of $\beta$ from this
value indicate that the quantum ring is essentially three-dimensional
and should not be treated as a bended quantum wire. As an example,
we present the parametrization of the quantum ring spectra for three
different shape configurations in Table~\ref{tab:mSpectra}. In the
case of a flat ring ($\eta_{0}=0.1,$) we see that all the calculated
states demonstrate the parabolic $m$ dependence, and this ring configuration
essentially follows the one-dimensional model. For the other two configurations,
however, the value of $\beta$ is measurably smaller than 2, and the
one-dimensional model of the ring does not describe the energy spectrum
of the "magnetic" excitations. In Fig.~\ref{fig:spectrumShapeMap}
we show a contour map for the magnetic spectrum shape parameter $\beta_{00}$
as a function of the ring shape parameters $\eta_{0}$ and $\sigma.$
The maps for higher $k$ and $q$ have similar structure. The map
demonstrates several interesting features.

First, there is a clear tendency for flat ring configurations ($\eta_{0}<0.125$)
to produce $m$-dependence of the spectrum very close to the textbook
$m^{2}$ behavior of 1D models. The configurations with more prominent
3D structure ($\eta_{0}>0.4$) generally produce spectra that deviate
from the 1D model quite substantially.

Second, the configurations of small $\sigma$ follow 1D-like dependence
on the magnetic quantum number $m$ for a broader range of configurations
independent, basically, of the parameter $\eta_{0}.$ This is not
surprising, as these configurations do look like as a 1D wire loop
for smaller $\eta_{0}$ and become flat as $\eta_{0}$ increases.

Finally, there is a special set of shapes about $\sigma\approx0.45$
and $\eta_{0}>0.95$ for which the deviation of the spectrum $m$-dependence
from the 1D model is the most prominent. It is interesting to note
that in the vicinity of this region we also see rapid change in the
behavior of the spectrum from purely parabolic to non-parabolic while
the variations of the ring shapes are rather small.

The use of oblate spheroidal coordinates for quantum ring calculations
allowed us to discover a nontrivial shape dependence of the ring
spectrum properties.

In this article we studied the model of a quantum ring formed in a
potential well with infinite walls. We have shown that the corresponding
Schr\"odinger equation admits separation of variables in oblate spheroidal
coordinates. This allowed us to construct a classification of single-particle
states in the quantum ring and to demonstrate essential dependence
of the single-particle spectrum on the shape of the quantum ring.
In particular, the most demonstrating example of such shape dependence
can be seen by studying the dependence of the spectrum on the magnetic
quantum number. We see, that a rather common assumption of parabolic
dependence of 1D model should not be taken for granted, and quantum
rings of many shapes that resemble realistic configurations are expected
to demonstrate the spectra that scale as $m^{\beta}$ with $1.3<\beta<2.$

Obviously, the model need further
refinements to be employed in quantitative studies.
The first step towards more elaborated models would rely on
the existence of potentials that admit separation of variables in
oblate spheroidal coordinates while reproducing 
the interaction of a particle with the nano-structure of
interest. Fortunately, there are known classes of potentials that
fit this description, and we are planning to study such potential
models in the nearest future. Studying the interactions of spheroidal
quantum rings with external fields that do not break the symmetry
of the system also seems an interesting direction of research.
\vskip -0.8cm
\section*{Acknowledgements}
\vskip -0.5cm
Authors want to thank Prof.~Slavyanov, Prof.~Verbin, Prof.~Abarenkov and Dr.~Kovalenko 
for encouraging discussions.

\end{document}